\documentclass[preprint,12pt,authoryear]{elsarticle}

\usepackage{amssymb}

\usepackage{graphicx}
\usepackage{dcolumn}
\usepackage{bm}
\usepackage{color}
\newcommand{\chM}{}

\journal{Physics Letters A}

\begin{document}

\begin{frontmatter}



\title{Information Theory Perspective on Network Robustness}


\author{Tiago A. Schieber}
 \address{%
Departmento de Engenharia de Produ\c c\~ao, \\Universidade Federal de Minas Gerais, Belo Horizonte, MG, Brazil
}
\address{%
Industrial and Systems Engineering, University of Florida, Gainesville, FL, USA
}%

\author{Laura Carpi}
\address{
Departament de F\'isica i Enginyeria Nuclear,\\
Universitat Polit\`ecnica de Catalunya.
Colom 11, Terrassa 08222, Barcelona, Spain
}%

\author{Alejandro C. Frery}
\address{
Laborat\'orio de Computa\c c\~ao Cient\'ifica e An\'alise Num\'erica (LaCCAN),\\
Universidade Federal de Alagoas, Macei\'o, Alagoas, Brazil
}%

\author{Osvaldo A Rosso}
\address{Instituto de F\'isica, Universidade Federal de Alagoas, Macei\'o, Alagoas, Brazil}%
\address{Instituto Tecnol\'ogico de Buenos Aires (ITBA), Ciudad Aut\'onoma de Buenos Aires, Argentina}

\author{Panos M. Pardalos}
\address{%
Industrial and Systems Engineering, University of Florida, Gainesville, FL, USA
}%

\author{Mart\'in G. Ravetti}
 \ead{martin.ravetti@dep.ufmg.br}
 \address{%
Departmento de Engenharia de Produ\c c\~ao, \\Universidade Federal de Minas Gerais, Belo Horizonte, MG, Brazil
}%
 \address{Departament de F\'isica Fonamental, Universitat de Barcelona, Barcelona, Spain}

\begin{abstract}

A crucial challenge in network theory is the study of the robustness of a network when facing a sequence of failures. In this work, we propose a dynamical definition of network robustness based on Information Theory, that considers measurements of the structural changes caused by failures of the network's components.
Failures are defined here, as a temporal process defined in a sequence. Robustness is then evaluated by measuring dissimilarities between topologies after each time step of the sequence, providing a dynamical information about the topological damage. We thoroughly analyze the efficiency of the method in capturing small perturbations by considering different probability distributions on networks. In particular, we find that distributions based on distances are more consistent in capturing network structural deviations, as better reflects the consequences of the failures. Theoretical examples and real networks are used to study the performance of this methodology. 
\end{abstract}


\end{frontmatter}


\section{\label{sec:Intro}Introduction}

There are several works dealing with the concept of robustness, however, there is still no consensus on a definitive definition. Robustness is usually described as the ability of the network to continue performing~\citep{Albert2000}, or, as the capacity in maintaining its functionality after failures or attacks~\citep{Fiedler1973,Iyer2013,Dekker2004}. These general definitions are perhaps the most used in the literature, however, they cannot exactly grasp the complexity of the concept that network's robustness could have. Some other works describe robustness as the capacity of the network in maintaining its efficiency in the presence of failures~\citep{Crucitti2003,Crucitti2004}. In some sense, this definition provides more information about the network's topological structure, as its efficiency depends on the network's shortest path lengths~\citep{Latora2007}.

The study of how robust a network is when facing random failures or targeted attacks is a major challenge in network theory. 
{\chM Several methodologies have been proposed to measure network robustness. Approaches based on information routing~\citep{boginski2009,puaa2012,pu2013}, structural controllability~\citep{liu2011,pu2012} or in the proposal of a more destructive attack strategy in networks~\citep{arulselvan2009,pua2015}, can be found in the literature; being the most popular those based on percolation theory~\citep{Albert2000,Cohen2000,Callaway2000}, and on the size of the biggest connected component (BC)~\citep{Holme2002,Allesina2009,Salathe2010,Iyer2013}. Although these measures showed to be useful in many cases, they are not as sensitive as they should, to the detection of failures that do not disconnect the network or that do not modify its diameter. 
Depending on the network structure, it is possible to attack great part of it, keeping these measures blind to the changes. 

In this work, we propose a measure for network robustness based on the Jensen-Shannon divergence, an Information Theory quantifier that already showed to be very effective in measuring small topological changes in a network~\citep{Carpi2011,Schieber2013,schieber2014}. This method considers failures occurring in a temporal sequence capturing, in some sense, the dynamics of the role of the remaining links after each single failure. The Jensen-Shannon divergence quantifies the topological damage of each time step due to failures, and the robustness measure provides the cumulative information of these sequential topological damages. It is worth noticing that this approach does not consider the consequences of the dynamical process operating through the network.}

\section{\label{sec:method}Methodology}

{\chM Quantification of network} robustness could be thought as the distance that a given topology is apart from itself after a failure. We assume that the robustness value ranges from 0, the {\chM greatest} variation, to 1, unchanged characteristics. In other words, a higher robustness value implies in smaller structural changes. In this work we consider {\chM a link failure, its removal, and a node failure in the removal of all it incident links}. 

Let $G$ be a network defined by a set  $V(G)$ of $N$ nodes, a set ${ \cal E}(G)$ of $M$ links {\chM and a set $W(E(G))$ containing the edges strengths}. A network failure event $f$ is defined as the removal of a subset of edges $f\subset{\cal E}(G)$.  A time-ordered sequence of failures  ${\cal F} =\{ f_{t_1}, f_{t_2},\dots,f_{t_n}\}$ in $G$ can be interpreted as a sequence of the resulting networks after each event $(G_{t_i})_{i\in\{0,\;1,\ldots, n \}}$ such that $G_{t_0}=G$ and $G_{t_i}$ is the network obtained after the failure $f_{t_i}$ in $G_{t_{i-1}}$. For simplicity, here we consider only discrete time intervals given by $t_i=i$. 

Considering the set ${\cal N}_{\cal F}$ of all possible sequences of failures in a network $G$, a {\em{robustness function} with respect to $G$ is a function $R \colon{\cal N}_{\cal F}\rightarrow [0,1]$}. The distance between two networks is computed as the distance between probability distributions used to characterize them. Without loss of generality, discrete distributions will be considered henceforth. 
 
The {\em Jensen-Shannon divergence} between two probability distributions $P$ and $Q$ is defined as the Shannon entropy of the average minus the average of the entropies. This measure was proven to be the {\em square of a metric} between probability distributions~\citep{Lin1991}, bounded by 1, and defined as:  

$$
{\cal J}^H(P,Q)=H\left(\frac{P+Q}{2}\right)-\frac{H(P)+H(Q)}{2} ,
$$ being {\chM $H(P)=-\sum_ip_i\log_2p_i$}, the entropy that measures the {\em amount of uncertainty} in a probability distribution. {\chM Readers are referred to the Supplementary Information material ({\bf SI}) for a discussion on the continuous case}.

It is possible then, to define the {\em robustness} of $G$, for any given sequence of $n$ failures $(G_t)_{t\in\{1,\;2,\dots, n \}}$ and probability distribution $P$ as:
\begin{equation}\label{eq:robustness}
R_P(G|(G_t)_{t\in\{1,\;2,\ldots, n \}})=\prod_{t=1}^{n}\left[1-{\cal J}^H(P(G_t),P(G_{t-1}))\right],
\end{equation} 
being $G_0=G$.

A more suitable form of equation (\ref{eq:robustness}) can be obtained via recurrence relation:
\begin{equation}\label{eq:robustness1}
R_P(G|(G_t)_{t\in\{1,\;2,\dots, n \}})=\prod_{t=1}^{n}R_P(G_{t-1}|G_t),
\end{equation} 
in which, for each time step, $R_P(G_{t-1}|G_t)$ indicates
how affected the topology of the network $G_{t-1}$ is after a single failure resulting in $G_t$. The robustness function depends on the network's topology, and also on the sequence of failures. {\chM The same link possesses different importance (effect) in the topology, depending on its position in the failure sequence. The use of the product of the temporal fluctuations  ($R_P(G_{t-1}|G_t)$) allows us to have a perception of {\chM the temporal damage due the sequence of failures}.}


It is important to notice that the computation of the network robustness, when defined as in equation (\ref{eq:robustness}) could consider any probability distribution able to represent features of the network. This work specifically considers the degree distribution, commonly used to characterize network's structures, and the distance distribution that contains rich information about the graph structure. The degree and distance distributions are here defined for unweighted and undirected networks. {\chM In Section I of the {\bf SI} readers can find a discussion other network robustness measurements and in section VI the analysis of a directed and weighted network is performed}.   

Given a node $i$, its {\em degree}, represented by $k_i$, is the number of edges incident to it. Then, the {\em degree distribution} $P_{deg}(k)$ is the fraction of nodes with degree $k$. The {\em distance} from the node $i$ to node $j$, $d_{i,j}$, is the length of the shortest path from $i$ to $j$. If there is no such path from $i$ to $j$, $d_{i,j}$ equals $\infty$. Then, the {\em distance distribution} $P_{\delta}(d)$ is the fraction of pairs of nodes at distance $d$. Both degree and distance distributions are discrete and {\chM defined on} the sets $\{0,\;1,\dots,\;N-1\}$ and  $\{1,\;2,\dots,\;N-1,\;\infty\}$, respectively.

\section{\label{sec:exper}Discussions and Computational Experiments}
With the aim of comparing the performance of the methodology based on $R_{P_{deg}}$ and $R_{P_\delta}$ with commonly used methods based on the biggest connected component ($R_{bc}$) and percolation ($R_{\pi_d}$), we consider the deletion of a single link on a complete graph with $N$ nodes, the most robust unweighted and undirected graph.

$R_{bc}$ is obtained by computing the fraction of nodes belonging to the biggest connected component.  In this case, $R_{bc}$ does not notice the removal of any link, as no disconnection is achieved. It is possible to strategically remove $N^2/2-2 N +1$ links, leaving just the minimum spanning tree, where robustness measures based on the biggest connected component remain blind to these attacks. 

Percolation based measures correlate the robustness value of the network with the critical percolation threshold and can be computed in several ways. One of the most common methods depends on the number of links removed until increasing the diameter of the network. $R_{\pi_d}$ indicates the variation of the original diameter $d_0$ with respect to diameter $d$ after a sequence of failures, computed by $R_{\pi_d}=d_0/d$.

{\chM In the case of the complete graph, the deletion of a single link increases the network diameter in one unit, however, after the first attack $R_{\pi_d}$ may become unable to detect subsequent events.  In order to increase the network's diameter in one more unit, the removal of $N-2$ specific links are needed.} 

{\chM The proposed robustness measure is able to detect the removal of any single link of the network, independently of which probability distribution ($P_{deg}$, $P_{\delta}$) is evaluated. Values for $R_{P_{deg}}$ and $R_{P_\delta}$ can be then, easily computed as functions of $N$.
The removal of a single link in a complete graph with $10^6$ nodes, implies in changes of the order of $10^{-6}$ for $R_{P_{deg}}$ and $10^{-13}$  for  $R_{P_\delta}$. For a complete graph of $N=10^7$, changes are of the order of $10^{-7}$ for $R_{P_{deg}}$ and $10^{-15}$ for $R_{P_\delta}$.}

 Among the measures here considered, only $R_{P_{deg}}$ and $R_{P_\delta}$ showed to be capable of capturing the removal of any single link, showing a gradual decrease in the robustness values, as more links are removed from the network.
This could be of relevance in situations in which it is necessary to plan the inclusion of new links to improve the robustness of the network. Methodologies based on the size of the biggest connected component, or on the percolation threshold, are not able to properly guide in this purpose. {\chM  It is important to point out,  that the robustness measure here proposed, depends not only on the network topology but on the sequence of failures over time, aiming to quantify the vulnerability of a given structure under a series of deterministic or stochastic failures. The process of fixing failures cannot be measured in the same way, but the degree and distance probability distributions seem to be adequate to this purpose.}
 
There are interesting differences between $P_{deg}$ and $P_{\delta}$. The computational complexity to obtain the degree distribution is linear, plus a constant cost to update it, after any link removal. The best known algorithm for obtaining $P_{\delta}$ requires ${\cal O}(N^{2.376})$ in time complexity~\citep{Tang2009}, and the computational cost of the PDF update depends on the link removed. However new algorithms as the ANF or HyperANF (algorithms based on HyperLogLog counters) offer an extremely fast and precise approach~\citep{Boldi2013,Palmer2002,Crescenzi2011,Boldi2011}, obtaining very good approximations of the distance {\chM probability distribution} for graphs with millions of nodes in a few seconds. {\chM In the {\bf SI} readers can find a table with the computational complexity of the most common methods.}

Another important comparison is the information that can be assessed from both distributions, and their correlation with topological structures. The network's average degree, mean degree and the minimum and maximum degree are immediately obtained from the degree distribution. The network's efficiency, diameter, average path length, fraction of disconnected pairs of nodes and other distance features are easily obtained from the distance distribution. 

\begin{figure}
\centering
\includegraphics[scale=0.8]{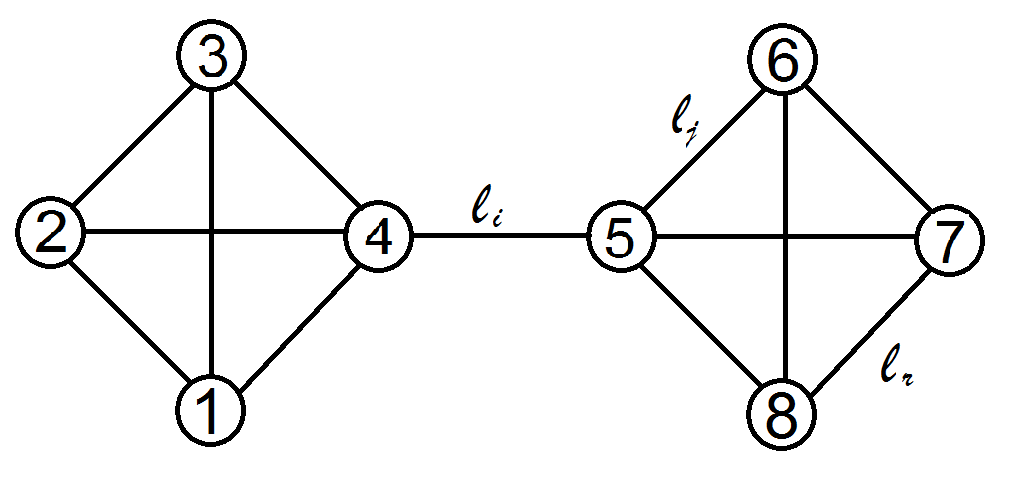}

\begin{tabular}{lcccc}
Edge Removed & $R_{P_\delta}$ & $R_{P_{deg}}$& $R_{bc}$& $R_{\pi_d}$\\
\hline
$\ell_i$ & 0.447 & 0.862&0.500& 0.000 \\
$\ell_j$ & 0.943 & 0.922&1.000& 0.750\\
$\ell_r$ & 0.998 & 0.857&1.000&1.000\\
\end{tabular}

\caption{\label{fig.comparacao} Computation of the structural robustness for three different {\chM single} edge removal: $\ell_i$, $\ell_j$ and $\ell_r$, respectively. }
\end{figure}

{\chM Figure~\ref{fig.comparacao} shows a simple network structure to analyze the correlation of the robustness values with different topological changes. Individual removal of links  $\ell_{i}$, $\ell_{j}$ and $\ell_{r}$ are performed. $R_{bc}$ only detects the disconnection of the biggest connected component, being not sensitive to the removal of $\ell_{j}$ and $\ell_{r}$.  $R_{\pi_d}$ detects the removal of $\ell_j$ and $\ell_{i}$, but fails in capturing the removal of $\ell_{r}$, as there is no modification in the diameter. The $R_{P_{deg}}$ detects every single failure, however, its value does not properly reflect the network's disconnection ($\ell_{i}$).} The measure based on the distance distribution ($R_{P_\delta}$) captures, in a more appropriate way, each of the above-mentioned network failures, especially those aspects concerning disconnections on a connected network. This example captures important advantages and disadvantages of each robustness measure. In the {\bf SI}, other quantifiers to measure robustness are also evaluated. However, the use of the distance distribution shows to be the most adequate for this analysis.

{\chM Let us now analyze two sequences of failures considering $\ell_i$ and $\ell_{j}$. If link $\ell_i$ fails at instant $t=1$ and link $\ell_j$ fails at instant $t=2$, $R_{P_{\delta}} = 0.4377$. Now, if the sequence is inverted considering link $\ell_{j}$ failing at instant $t=1$ and link $\ell_{i}$ failing at instant $t=2$, $R_{P_{\delta}}$= $0.4564$. This example depicts how the roles and topological importance of the remaining links after a failure are reflected by $R_{P_{\delta}}$}.

{\chM We test the proposed methodology on several real networks, nevertheless, only the results for two of them are depicted in the main text}, the Dolphin Social Network~\citep{Lusseau2003} and the Western States Power Grid of the United States network~\citep{Watts1998}. {\chM Readers are referred to the {\bf SI} section V for applications on other networks}.

The Dolphin network is an undirected social network of bottlenose dolphins ({\it genus Tursiops}). The nodes are the bottlenose dolphins  of a  community from New Zealand, where an edge indicates a frequent association {\chM between dolphin pairs occurring more often than expected by
chance~\citep{Lusseau2003}}. The dolphins were observed between 1994 and 2001. It presents $N=62$, $M=159$, an average degree of $5.13$, an average path length of $3.357$, and a clustering coefficient of $0.258$. 

The Power Grid Network is the undirected and unweighted representation of the topology of the Western States Power Grid of the United States, compiled by Duncan Watts and Steven Strogatz~\citep{Watts1998}. It presents, $N=4941$, $M=6594$, an average degree of $2.67$, an average path length of $18.99$, and a clustering coefficient of $0.103$.

\begin{figure*}[htb!]
 \includegraphics[bb=0 0 510 416,width=\linewidth]{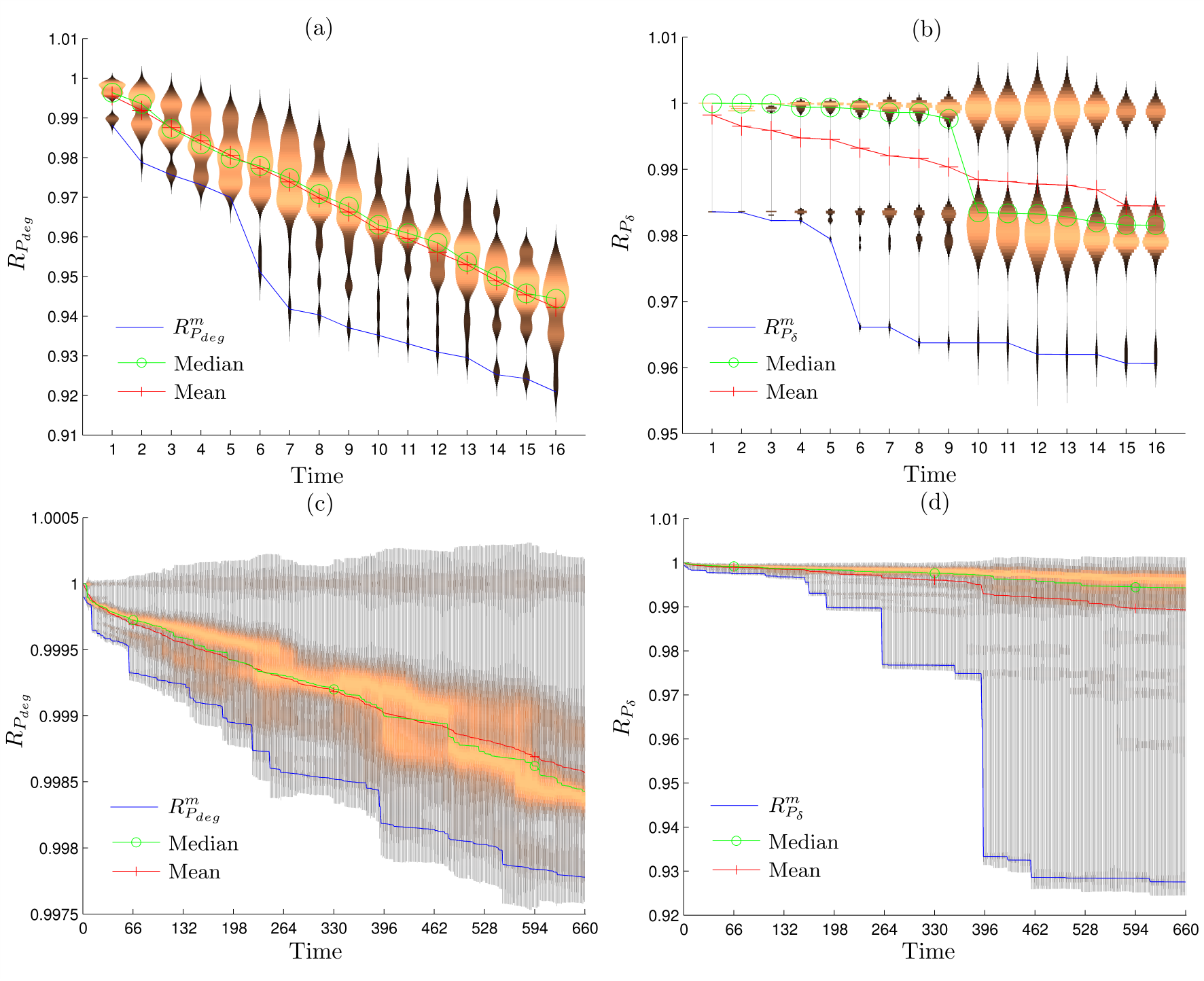}
 \caption{\label{fig.combinacao}Robustness measures under random failures for Dolphin and Power Grid networks. At each time step, a random edge is disconnected from the network and $R_{P_{deg}}$, $R_{P_\delta}$ functions are computed. The experiment is independently executed $30$ times. Results for the Dolphin networks are depicted in (a) $R_{P{deg}}$ and (b) $R_{P_{\delta}}$. Results for the Power Grid network are presented in (c) $R_{P_{deg}}$ and (d) $R_{P_\delta}$. In all cases, at each time step, the minimum robustness values are also indicated by $R^m_{P}$.}
\end{figure*}

{\chM In both cases, at each time step a single link is randomly removed until the global disconnection of approximately 10\% of their links. Thirty independent experiments were performed and, at each time step, the robustness measure for each experiment is computed.  Figure \ref{fig.combinacao} depicts the  composition of violin plots of the robustness value, where  $R^m_{P}$ indicates the minimum robustness value found at each time step. }

It is possible to see from Figures (\ref{fig.combinacao}.a) and (\ref{fig.combinacao}.c) that the robustness measure computed from the degree distribution shows a smoother behavior, as it is unable to detect {\chM cluster} disconnections. This is not the case for the distance distribution, in which the fraction of disconnected pairs of nodes is detected (see Figures (\ref{fig.combinacao}.b) and (\ref{fig.combinacao}.d)). The large decrease in the $R_{P_\delta}$ values usually represents cluster disconnections from the network. As we are analyzing average values, the disconnection may occur in a fraction of the thirty independent experiments only. 

{\chM The large variability of single robustness values for the Dolphin network reflects the extent of the damage that certain failures can cause, showing the Dolphin network more susceptible to random failures than the US Power Grid. The robustness measures, in particular those computed through the distance distributions, Figures (\ref{fig.combinacao}.b) and (\ref{fig.combinacao}.d) also show big leaps when the link removal is around 6\% and 9\% for Dolphin and 3\% and 6\% for the Power Grid, indicating network's disconnections.}

\begin{figure*}[htb!]
 \includegraphics[bb=0 0 506 211,,width=\linewidth]{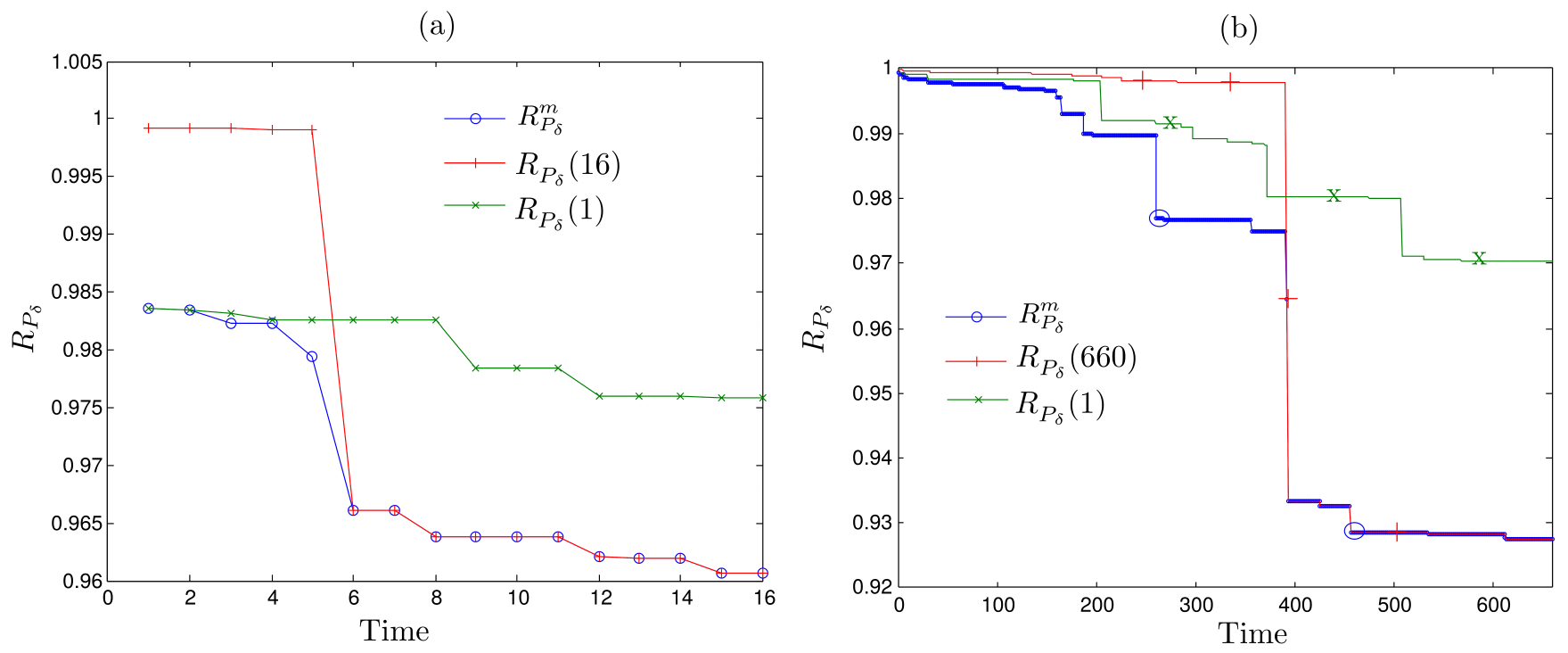}
 \caption{\label{figataques}Evolution of the $R_{P_\delta}$ of two different sequences of failures: the sequence that ends with the lowest robustness value, $R_{P_\delta}(16)$ and $R_{P_\delta}(660)$, and the sequence in which the first removal is most effective, $R_{P_\delta}(1)$. In both cases, at each time step, the minimum robustness values are also indicated by $R^m_{P_\delta}$. (a) Dolphin network and (b) US Power Grid network.}
\end{figure*}

{\chM Figure \ref{figataques} compares the $R^m_{P}$ values with two sequences of failures for each experiment; the sequences presenting the lowest  robustness value at the end of the attack ($R_{P_\delta}(16)$ for Dolphin and $R_{P_\delta}(660)$ for Power Grid) and the sequences with the lowest robustness value at the first time step ($R_{P_\delta}(1)$). Note that the sequences of failures resulting in lower robustness values are not the most efficient in destroying the network at the beginning of the process. This behavior occurs because the robustness measure provides cumulative information about the evolution of the state of the network (see Equation (\ref{eq:robustness1})). A small $R_P(G_t|G_{t-1})$ value indicates that, at time $t$, the failure of certain links is critical, generating bigger changes in the topology.}

\begin{figure*}[b!]
\centering
\includegraphics[width=\linewidth]{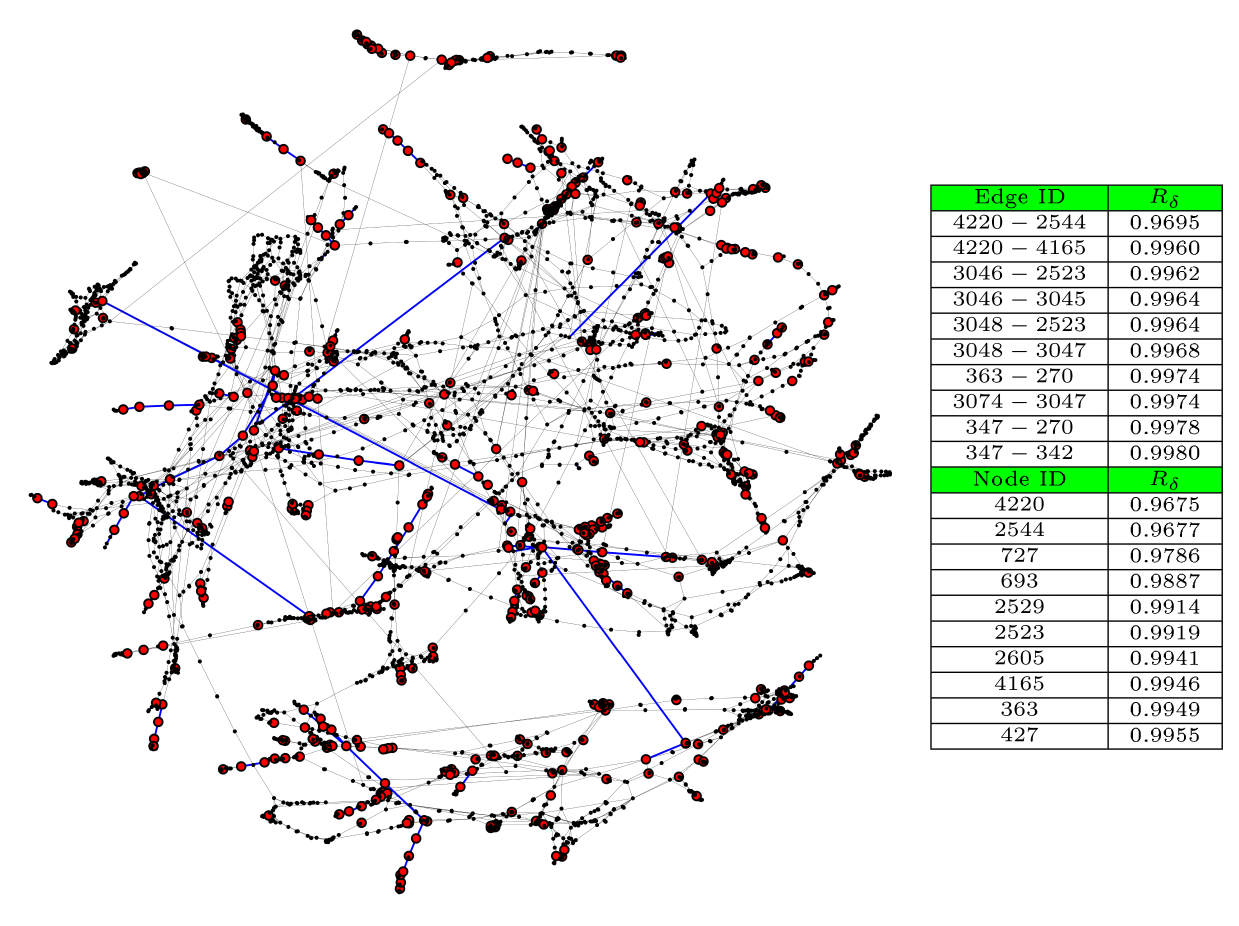}
\caption{\label{fig.powergrid_critical} Detection of the ten percent of the most critical links and nodes, considering $R_{P_\delta}$ over the Western States Power Grid of the United States network. Red nodes represent the fraction of 10 of network's vertices such that its single disconnection causes a big reduction on the $R_{P_\delta}$ value. Blue links represent the fraction of 10 of network's edges such that its single disconnection causes a big reduction on the $R_{P_\delta}$ value. The table shows the robustness values of the top 10 critical network elements. }
\end{figure*}

This methodology could also be applied to detect critical elements, such as the nodes and links in the US power grid network that, when individually removed, cause a major disturbance in the network's structure. Figure~\ref{fig.powergrid_critical} shows the 10 most critical links and nodes that produce the largest robustness values variation (see table in Figure~\ref{fig.powergrid_critical}). 
Critical elements for US power grid network identified by $R_{P_{deg}}$,  as well as results for Dolphin network, can be found in {\bf SI}; {\it cf.} Figures~S3-S5.  

{\chM The knowledge of critical elements  is of great importance to plan strategies either to protect or to efficiently attack networks.  In both scenarios, the knowledge about how the network continues to perform after failures is of paramount importance. It is interesting noticing that the problem of finding the best sequence of links to destroy the network can be solved through combinatorial optimization approaches. Readers are referred to {\bf SI}, section VI for a computational experiment considering targeted attacks in two directed and weighted real networks.}

\section{Final Remarks}
{\chM We propose a novel methodology to measure the robustness of a network to component failures or targeted attacks. This mathematical formulation is based on the consideration that the network robustness is a measure related to the distance that a given topology is apart from itself after a sequence of failures, rather that a single characteristic of the topology. This sequence is defined as a time dependent process in which, a subset of links is disconnected at each time step. The method provides a dynamic robustness profile that shows the response of the network's topology to each event, quantifying the vulnerability of these intermediate topologies. 

Although the methodology is comprehensive enough to be used with different probability distributions, the use of distances shows to be more consistent in capturing network structural deviations, in the sense that their values are correlated with the consequences of the failures in the network topology. Different from the methods found in the literature, the method can efficiently work with disconnections, as the distance PDF is able to acknowledge the fraction of disconnected pairs of nodes. Furthermore, it is able to detect all changes, including those perceived by $R_{bc}$ and $R_{\pi_d}$, resulting in a more general approach.}


\section{Acknowledgments}
Research partially supported by FAPEMIG, CNPq and grant RSF 14-41-00039. OAR acknowledges support from  CONICET.




\clearpage
\bibliographystyle{elsarticle-harv}

\bibliography{references_robustness_aps}

\begin{thebibliography}{30}
\expandafter\ifx\csname natexlab\endcsname\relax\def\natexlab#1{#1}\fi
\expandafter\ifx\csname url\endcsname\relax
  \def\url#1{\texttt{#1}}\fi
\expandafter\ifx\csname urlprefix\endcsname\relax\def\urlprefix{URL }\fi

\bibitem[{Albert et~al.(2000)Albert, Jeong, and Barab\'asi}]{Albert2000}
Albert, R., Jeong, H., Barab\'asi, A.-L., Jul. 2000. Error and attack tolerance
  of complex networks. Nature 406~(6794), 378--382.

\bibitem[{Allesina and Pascual(2009)}]{Allesina2009}
Allesina, S., Pascual, M., Sep. 2009. Googling food webs: Can an eigenvector
  measure species' importance for coextinctions? PLoS Comput Biol 5~(9),
  e1000494--.

\bibitem[{Arulselvan et~al.(2009)Arulselvan, Commander, Elefteriadou, and
  Pardalos}]{arulselvan2009}
Arulselvan, A., Commander, C.~W., Elefteriadou, L., Pardalos, P.~M., jul 2009.
  Detecting critical nodes in sparse graphs. Comput. Oper. Res. 36~(7),
  2193--2200.

\bibitem[{Boginski et~al.(2009)Boginski, Commander, and Turko}]{boginski2009}
Boginski, V., Commander, C.~W., Turko, T., 2009. Polynomial-time identification
  of robust network flows under uncertain arc failures. Optimization Letters
  3~(3), 461--473.

\bibitem[{Boldi et~al.(2011)Boldi, Rosa, and Vigna}]{Boldi2011}
Boldi, P., Rosa, M., Vigna, S., 2011. Robustness of social networks:
  Comparative results based on distance distributions. In: Proceedings of the
  Third International Conference on Social Informatics. SocInfo'11.
  Springer-Verlag, Berlin, Heidelberg, pp. 8--21.

\bibitem[{Boldi and Vigna(2013)}]{Boldi2013}
Boldi, P., Vigna, S., 2013. In-core computation of geometric centralities with
  hyperball: A hundred billion nodes and beyond. CoRR abs/1308.2144.

\bibitem[{Callaway et~al.(2000)Callaway, Newman, Strogatz, and
  Watts}]{Callaway2000}
Callaway, D.~S., Newman, M. E.~J., Strogatz, S.~H., Watts, D.~J., Dec 2000.
  Network robustness and fragility: Percolation on random graphs. Phys. Rev.
  Lett. 85, 5468--5471.

\bibitem[{Carpi et~al.(2011)Carpi, Rosso, Saco, and Ravetti}]{Carpi2011}
Carpi, L.~C., Rosso, O.~A., Saco, P.~M., Ravetti, M., Jan. 2011. Analyzing
  complex networks evolution through information theory quantifiers. Physics
  Letters A 375~(4), 801--804.

\bibitem[{Cohen et~al.(2000)Cohen, Erez, ben Avraham, and Havlin}]{Cohen2000}
Cohen, R., Erez, K., ben Avraham, D., Havlin, S., Nov 2000. Resilience of the
  internet to random breakdowns. Phys. Rev. Lett. 85, 4626--4628.

\bibitem[{Crescenzi et~al.(2011)Crescenzi, Grossi, Lanzi, and
  Marino}]{Crescenzi2011}
Crescenzi, P., Grossi, R., Lanzi, L., Marino, A., 2011. A comparison of three
  algorithms for approximating the distance distribution in real-world graphs.
  In: Marchetti-Spaccamela, A., Segal, M. (Eds.), Lecture Notes in Computer
  Science. Vol. 6595. Springer Berlin Heidelberg, pp. 92--103--.

\bibitem[{Crucitti et~al.(2003)Crucitti, Latora, Marchiori, and
  Rapisarda}]{Crucitti2003}
Crucitti, P., Latora, V., Marchiori, M., Rapisarda, A., Mar 2003. Efficiency of
  scale-free networks: error and attack tolerance. Physica A: Statistical
  Mechanics and its Applications 320, 622--642.

\bibitem[{Crucitti et~al.(2004)Crucitti, Latora, Marchiori, and
  Rapisarda}]{Crucitti2004}
Crucitti, P., Latora, V., Marchiori, M., Rapisarda, A., 2004. Error and attack
  tolerance of complex networks. Physica A: Statistical Mechanics and its
  Applications 340~(1–3), 388 -- 394, news and Expectations in
  Thermostatistics.

\bibitem[{Dekker and Colbert(2004)}]{Dekker2004}
Dekker, A.~H., Colbert, B.~D., 2004. Network robustness and graph topology. In:
  Proceedings of the 27th Australasian Conference on Computer Science - Volume
  26. ACSC '04. Australian Computer Society, Inc., Darlinghurst, Australia,
  Australia, pp. 359--368.
\newline\urlprefix\url{http://dl.acm.org/citation.cfm?id=979922.979965}

\bibitem[{Fiedler(1973)}]{Fiedler1973}
Fiedler, M., 1973. {Algebraic connectivity of graphs}. Czechoslovak
  Mathematical Journal 23~(98), 298--305.

\bibitem[{Holme et~al.(2002)Holme, Kim, Yoon, and Han}]{Holme2002}
Holme, P., Kim, B.~J., Yoon, C.~N., Han, S.~K., May 2002. Attack vulnerability
  of complex networks. Phys. Rev. E 65, 056109.

\bibitem[{Iyer et~al.(2013)Iyer, Killingback, Sundaram, and Wang}]{Iyer2013}
Iyer, S., Killingback, T., Sundaram, B., Wang, Z., Apr. 2013. Attack robustness
  and centrality of complex networks. PLoS ONE 8~(4), e59613--.

\bibitem[{Latora and Marchiori(2007)}]{Latora2007}
Latora, V., Marchiori, M., 2007. A measure of centrality based on network
  efficiency. New J. Phys. 9~(188).

\bibitem[{Lin(1991)}]{Lin1991}
Lin, J., Jan 1991. Divergence measures based on the shannon entropy.
  Information Theory, IEEE Transactions on 37~(1), 145--151.

\bibitem[{Liu et~al.(2011)Liu, Slotine, and Barabasi}]{liu2011}
Liu, Y.-Y., Slotine, J.-J., Barabasi, A.-L., May 2011. {Controllability of
  complex networks}. Nature 473~(7346), 167--173.
\newline\urlprefix\url{http://dx.doi.org/10.1038/nature10011}

\bibitem[{Lusseau et~al.(2003)Lusseau, Schneider, Boisseau, Haase, Slooten, and
  Dawson}]{Lusseau2003}
Lusseau, D., Schneider, K., Boisseau, O., Haase, P., Slooten, E., Dawson, S.,
  2003. {The bottlenose dolphin community of Doubtful Sound features a large
  proportion of long-lasting associations}. Behavioral Ecology and Sociobiology
  54~(4), 396--405.

\bibitem[{Palmer et~al.(2002)Palmer, Gibbons, and Faloutsos}]{Palmer2002}
Palmer, C.~R., Gibbons, P.~B., Faloutsos, C., 2002. Anf: A fast and scalable
  tool for data mining in massive graphs. In: Proceedings of the Eighth ACM
  SIGKDD International Conference on Knowledge Discovery and Data Mining. KDD
  '02. ACM, New York, NY, USA, pp. 81--90.

\bibitem[{Pu et~al.(2015)Pu, Lia, Michaelsonb, and Yanga}]{pua2015}
Pu, C.-L., Lia, S., Michaelsonb, A., Yanga, J., August 2015. Iterative path
  attacks on networks. Physics Letters A 379, 1633–1638.

\bibitem[{Pu et~al.(2012{\natexlab{a}})Pu, Pei, and Michaelson}]{pu2012}
Pu, C.-L., Pei, W.-J., Michaelson, A., 2012{\natexlab{a}}. Robustness analysis
  of network controllability. Physica A 391, 4420--4425.

\bibitem[{Pu et~al.(2012{\natexlab{b}})Pu, Si-YuanZhou, Wang, Zhang, and
  Pei}]{puaa2012}
Pu, C.-L., Si-YuanZhou, Wang, K., Zhang, Y.-F., Pei, W.-J., 2012{\natexlab{b}}.
  Efficient and robust routing on scale-free networks. PhysicaA 391, 866--871.

\bibitem[{Pu et~al.(2013)Pu, Yang, Pei, Tao, and Lan}]{pu2013}
Pu, C.-L., Yang, J., Pei, W.-J., Tao, Y.-T., Lan, S.-H., August 2013.
  Robustness analysis of static routing on networks. Physica A: Statistical
  Mechanics and its Applications 392~(15), 3293–3300.

\bibitem[{Salath{\'e} et~al.(2010)Salath{\'e}, Kazandjieva, Lee, Levis,
  Feldman, and Jones}]{Salathe2010}
Salath{\'e}, M., Kazandjieva, M., Lee, J.~W., Levis, P., Feldman, M.~W., Jones,
  J.~H., 12 2010. A high-resolution human contact network for infectious
  disease transmission. Proceedings of the National Academy of Sciences
  107~(51), 22020--22025.

\bibitem[{Schieber et~al.(2014)Schieber, Carpi, and Ravetti}]{schieber2014}
Schieber, T., Carpi, L., Ravetti, M., 2014. Evaluation of the copycat model for
  predicting complex network growth. In: Vogiatzis, C., Walteros, J.~L.,
  Pardalos, P.~M. (Eds.), Dynamics of Information Systems. Vol. 105 of Springer
  Proceedings in Mathematics Statistics. Springer International Publishing, pp.
  91--108.

\bibitem[{Schieber and Ravetti(2013)}]{Schieber2013}
Schieber, T.~A., Ravetti, M.~G., Dec. 2013. Simulating the dynamics of
  scale-free networks via optimization. PLoS ONE 8~(12), e80783--.

\bibitem[{Tang et~al.(2009)Tang, Wang, Wang, and Wei}]{Tang2009}
Tang, J., Wang, T., Wang, J., Wei, D., 2009. Efficient social network
  approximate analysis on blogosphere based on network structure
  characteristics. In: Proceedings of the 3rd Workshop on Social Network Mining
  and Analysis. SNA-KDD '09. ACM, New York, NY, USA, pp. 7:1--7:8.

\bibitem[{Watts and Strogatz(1998)}]{Watts1998}
Watts, D.~J., Strogatz, S.~H., 06 1998. Collective dynamics of /`small-world/'
  networks. Nature 393~(6684), 440--442.

\end{thebibliography}

\end{document}